\documentclass[11pt]{article} 
\usepackage[paperheight=9.44in,paperwidth=6.69in,left=1.85cm,right=1.8cm,bottom=2.5cm,top=3cm,twoside]{geometry}

\usepackage[linesnumbered,ruled,vlined,boxed]{algorithm2e}
\usepackage{authblk} 
\usepackage{fancyhdr} 
\usepackage{lastpage} 
\usepackage{textcomp} 
\usepackage{amsmath} 
\usepackage{amssymb} 
\usepackage{amsthm} 
\usepackage{graphicx} 
\usepackage{multirow} 
\usepackage{hyperref} 

\newcommand{\linia}{\noindent\rule{\linewidth}{0.5mm}\hrulefill} 

\SetAlFnt{\small}
\SetAlCapFnt{\small}
\SetAlCapNameFnt{\small}
\SetAlCapHSkip{0pt}
\IncMargin{-\parindent}

\usepackage{times}
\usepackage{titlesec}

\titleformat*{\section}{\large\bfseries}
\titleformat*{\subsection}{\normalsize\bfseries}

\fancypagestyle{firststyle}
{
\fancyhf{}
\fancyfoot{}
\fancyhead{}
\fancyhead[CE]{\upshape Zeszyty Naukowe WWSI, No 13, Vol. 9, 2015, pp. \thepage-\pageref{LastPage}}
\fancyhead[CO]{\upshape Zeszyty Naukowe WWSI, No 13, Vol. 9, 2015, pp. \thepage-\pageref{LastPage}}
\fancyfoot[L]{\footnotesize\bfseries Manuscript received April 11, 2015}
\fancyfoot[LE,RO]{\thepage}
}

\title{\large \bfseries Behavioral modeling of stressed MOSFET} 
\author{\normalsize Zenon Gniazdowski\thanks{E-mail: zgniazdowski@wwsi.edu.pl}}
\affil{\normalsize Warsaw School of Computer Science}

\date{\vspace{-5ex}}
\providecommand{\keywords}[1]{\textbf{\textit{Keywords ---}} #1}

\begin{document}

\maketitle 
\thispagestyle{firststyle} 

\linia
\begin{abstract}
 \noindent In this paper piezoconductivity phenomenon in MOSFET channel is discussed and extension of drain current model with possibility of stress consideration is proposed. Analysis of obtained model combined with examination of stress components inherent in the MOSFET channel as well as distributions of specific piezoconductance coefficients on a plane of channel can show which directions of transistor channel are desirable for improvement of MOSFET performances. This model gives possibility to predict optimal transistor channel orientation, for the given stress state in MOSFET channel. Possible simplification of this model is considered. In particular, stress state and significant piezoconductance coefficient distributions on planes $\{100\}$, $\{110\}$ as well as $\{111\}$ are analyzed. For assumed particular cases of stress state in the channel, final models of MOSFT for considered specific planes are given.
\end{abstract}

\section{Introduction}\label{Section:Introduction}
When mechanical load is applied, anisotropic change of conductivity is observed on the silicon structure. This effect can be employed for improvement of MOSFET performances. In this paper, behavioral description of MOSFET drain current, under low mechanical load, from linear piezoconductivity point of view will be considered. For this purpose, it is assumed that description of piezoconductivity for bulk layers is also valid for inversion layer in MOSFET channel. Based on this assumption general model of stressed MOSFET will be derived as well as possible simplifications will be considered.

\section{Preliminaries}
Piezoconductivity is an anisotropic phenomenon. Anisotropy can be described with the use of tensor calculus. In mathematical description of piezoconductivity, vector notation of tensors rank two (conductivity, stress) and matrix notation of tensors rank four (piezoconductivity) will be used in this article \cite{CSS58}, instead of tensor notation.

\subsection{Anisotropy of Ohm{\textquotesingle}s law}
\noindent In conductive layer the relation between density current $j={\left[j_1,j_2,j_3\right]}^T$ and electric field $E={\left[E_1,E_2,E_3\right]}^T$ can be described as follows \cite{CSS58} \cite{JFN57}:

\begin{equation}\label{Eq1}
\begin{bmatrix}
j_1 \\ 
j_2 \\ 
j_3 \end{bmatrix}
{=}
\begin{bmatrix}
\kappa_{11} & \kappa_{12} & \kappa_{13} \\ 
\kappa_{21} & \kappa_{22} & \kappa_{23} \\ 
\kappa_{31} & \kappa_{32} & \kappa_{33} \end{bmatrix}
\cdot \begin{bmatrix}
E_1 \\ 
E_2 \\ 
E_3 \end{bmatrix}.
\end{equation}

\noindent In (\ref{Eq1}) square matrix represents conductivity, which is second rank tensor with nine components \textit{$\kappa$${}_{ij}$}. This tensor (denoted as $K$) is symmetrical according to change of indexes \cite{CSS58} \cite{JFN57} \cite{Now95}. It means that $K$ has only six independent components:

\begin{equation}\label{Eq2}
K= \begin{bmatrix}
\kappa_{11} & \kappa_{12} & \kappa_{13} \\ 
\kappa_{21} & \kappa_{22} & \kappa_{23} \\ 
\kappa_{31} & \kappa_{32} & \kappa_{33} \end{bmatrix}=
\begin{bmatrix}
\kappa_{1} & \kappa_{6} & \kappa_{5} \\ 
\kappa_{6} & \kappa_{2} & \kappa_{4} \\ 
\kappa_{5} & \kappa_{4} & \kappa_{3} \end{bmatrix}.
\end{equation}

\noindent It can be represented as a vector with six components:

\begin{equation}\label{Eq3}
{K}{=}\begin{bmatrix}\kappa_{1},\kappa_{2},\kappa_{3},\kappa_{4},\kappa_{5},\kappa_{6}\end{bmatrix}^T. 
\end{equation}

\noindent For isotropic case, both vectors \textit{E} and \textit{j} are parallel. From here, \textit{$\kappa$${}_{1}$=$\kappa$${}_{2}$=$\kappa$${}_{3}$}=\textit{$\kappa$} and \textit{$\kappa$${}_{4}$=$\kappa$${}_{5}$=$\kappa$${}_{6}$=0}. Hence, in vector notation isotropic conductivity is a vector as follows:
\begin{equation}\label{Eq4}
K=
\begin{bmatrix}
\kappa,\kappa,\kappa,0,0,0
\end{bmatrix}
^T. 
\end{equation}

The state of stress in elastic body can be described as a second order symmetrical tensor, with only six independent components \cite{JFN57}. It can be represented also as a vector:

\begin{equation}\label{Eq5}
\sigma=
\begin{bmatrix}
\sigma _{1},\sigma _{2},\sigma _{3},\sigma _{4},\sigma _{5},\sigma _{6}
\end{bmatrix}^T.
\end{equation}
\noindent Components \textit{$\sigma$${}_{1}$, $\sigma$${}_{2}$, $\sigma$${}_{3}$ }represent normal stresses and components \textit{$\sigma$${}_{4}$, $\sigma$${}_{5}$, $\sigma$${}_{6}$} are shear stresses. Without stress, the conductivity of silicon crystal is isotropic. When the stress is applied, anisotropic change of conductivity is observed on the structure:
\begin{equation}\label{Eq6}
\Delta {K}{=}
\begin{bmatrix}
\Delta \kappa_{1},\Delta \kappa_{2},\Delta \kappa_{3},\Delta \kappa_{4},\Delta \kappa_{5},\Delta \kappa_{6}
\end{bmatrix}^T.
\end{equation}

\noindent In this case effective conductivity is a sum of isotropic conductivity (\ref{Eq4}) and anisotropic change of conductivity (\ref{Eq6}):
\begin{equation}\label{Eq7}
{K}{=}{{K}}_0{+}\Delta {K}.
\end{equation}
\noindent Relative change of conductivity can be described as follows:

\begin{equation}\label{Eq8}
\frac{\Delta K}{\kappa }{=}\frac{1}{\kappa}\cdot 
 \begin{bmatrix}
{\Delta \kappa}_{1} \\ 
{\Delta \kappa}_{2} \\ 
{\Delta \kappa}_{3} \\ 
{\Delta \kappa}_{4} \\ 
{\Delta \kappa}_{5} \\ 
{\Delta \kappa}_{6} \end{bmatrix}
{=} \begin{bmatrix}
{\Pi }_{11} & {\Pi }_{12} & {\Pi }_{13} & {\Pi }_{14} & {\Pi }_{15} & {\Pi }_{16} \\
{\Pi }_{21} & {\Pi }_{22} & {\Pi }_{23} & {\Pi }_{24} & {\Pi }_{25} & {\Pi }_{26} \\
{\Pi }_{31} & {\Pi }_{32} & {\Pi }_{33} & {\Pi }_{34} & {\Pi }_{35} & {\Pi }_{36} \\ 
{\Pi }_{41} & {\Pi }_{42} & {\Pi }_{43} & {\Pi }_{44} & {\Pi }_{45} & {\Pi }_{46} \\ 
{\Pi }_{51} & {\Pi }_{52} & {\Pi }_{53} & {\Pi }_{54} & {\Pi }_{55} & {\Pi }_{56} \\ 
{\Pi }_{61} & {\Pi }_{62} & {\Pi }_{63} & {\Pi }_{64} & {\Pi }_{65} & {\Pi }_{66}
\end{bmatrix}
\cdot 
\begin{bmatrix}
\sigma_{1} \\ 
\sigma_{2} \\ 
\sigma_{3} \\ 
\sigma_{4} \\ 
\sigma_{5} \\ 
\sigma_{6} 
\end{bmatrix}
{=}\Pi \sigma.
\end{equation}

\noindent Matrix $\Pi$ in (\ref{Eq8}) represents piezoconductivity tensor. Between changes in resistivity and conductivity, there is a relation \cite{www}:
\begin{equation}\label{Eq9}
\frac{\Delta \rho}{\rho}{=-}\frac{\Delta {K}}{\kappa}.
\end{equation}
\noindent It means that there is equivalence between piezoconductivity and piezoresistivity matrices in a sense of opposition:
\begin{equation}\label{Eq10}
\Pi {=-}\pi.
\end{equation}
\subsection{Piezoconductivity on arbitrary coordinate system}
It is assumed that there is known piezoconductivity matrix in coordinate system that is conformable to elementary cell of silicon. For finding distribution of piezoconductance coefficients on arbitrary chosen coordinate system it is required square matrix $a$, of dimension $3\times3$ \cite{JFN57}, which describes transformation from old coordinate system to the new one:
\begin{equation}\label{Eq11}
a{=} \begin{bmatrix}
a_{11} & a_{12} & a_{13} \\ 
a_{21} & a_{22} & a_{23} \\ 
a_{31} & a_{32} & a_{33} \end{bmatrix}
\buildrel{def}\over= \begin{bmatrix}
l_{1} & m_{1} & n_{1} \\ 
l_{2} & m_{2} & n_{2} \\ 
l_{3} & m_{3} & n_{3} \end{bmatrix}.
\end{equation}
\noindent From here, matrix $\alpha $ can be calculated \cite{CSS58}:
\begin {equation}\label {Eq12} 
\alpha {=} \begin {pmatrix} 
l^ {2} _ {1} & m^ {2} _ {1} & n^ {2} _ {1} & {2} m_ {1} n_ {1} & {2} n_ {1} l_ {1} & {2} l_ {1} m_ {1} \\
l^ {2} _ {2} & m^ {2} _ {2} & n^ {2} _ {2} & {2} m_ {2} n_ {2} & {2} n_ {2} l_ {2} & {2} l_ {2} m_ {2} \\
l^ {2} _ {3} & m^ {2} _ {3} & n^ {2} _ {3} & {2} m_ {3} n_ {3} & {2} n_ {3} l_ {3} & {2} l_ {3} m_ {3} \\
l_ {2} l_ {3} & m_ {2} m_ {3} & n_ {2} n_ {3} & m_ {2} n_ {3} {+} n_ {2} m_ {3} & l_ {2} n_ {3} {+} n_ {2} l_ {3} & l_ {2} m_ {3} {+} m_ {2} l_ {3} \\
l_ {3} l_ {1} & m_ {3} m_ {1} & n_ {3} n_ {1} & m_ {1} n_ {3} {+} n_ {1} m_ {3} & l_ {1} n_ {3} {+} n_ {1} l_ {3} & l_ {1} m_ {3} {+} m_ {1} l_ {3} \\
l_ {1} l_ {2} & m_ {1} m_ {2} & n_ {1} n_ {2} & m_ {1} n_ {2} {+} n_ {1} m_ {2} & l_ {1} n_ {2} {+} n_ {1} l_ {2} & l_ {1} m_ {2} {+} m_ {1} l_ {2}
\end {pmatrix} 
\end {equation} 

\noindent This matrix is needed for piezoconductance coefficients $\Pi \textquotesingle$ calculation in a new coordinate system:
\begin{equation}\label{Eq13}
{\Pi }{{'}}{=}\alpha\Pi \alpha^{{-}{1}}{=-}\alpha\pi \alpha^{{-1}}.
\end{equation}

\noindent Results of transformation (\ref{Eq13}) can be presented in polar coordinate system \cite{Kan82} or in Cartesian coordinate system. This paper results will be presented with the use of Cartesian coordinate system only. For assumed planes $\{100\}$, $\{110\}$ and $\{111\}$, given in \cite{Gni05} method will be used for finding matrix (\ref{Eq11}).

\subsubsection{Transformation matrix on plane $\{100\}$}
\begin{figure}
\centering
\includegraphics[width=10cm]{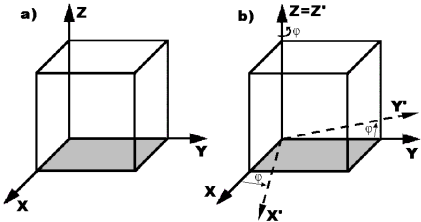}
\caption{Piezoconductance on a plane $\{100\}$: a) the plane $(001)$; b) rotation of coordinate system around $Z$-axis.}\label{fig1}
\end{figure}

Cartesian coordinate system conformed to crystallographic axes of silicon is considered. It is assumed that on plane $(001)$ current flows along $X$-axis. For investigation of piezoconductivity coefficients behavior on this plane, it is needed rotation of coordinate system located on $(001)$ around $Z$-axis by angle $\varphi$ (Fig. \ref{fig1}). Angles between coordinate system axes before rotation and after rotation are described in Table \ref{Tab1}. Therefore, specific matrix (\ref{Eq11}) which includes cosines of above-mentioned angles has a form \cite{JFN57}:
\begin{equation}\label{Eq14}
a{=}R(Z,\varphi){=}
\begin{bmatrix}
{\cos \varphi} & {\sin \varphi} & 0 \\ 
{-}{\sin \varphi\ } & {\cos \varphi\ } & 0 \\ 
0 & 0 & 1 \end{bmatrix}
\end{equation}

\noindent On plane $(001)$ range $0\leq\varphi\leq90^0$ of rotation angle is enough to describe behavior of piezoconductance coefficients, because of silicon cell symmetry. Based on formula (\ref{Eq13}), distribution of piezoconductance coefficients on plane $\{100\}$ can be received.

\begin{table}
\centering
\caption{Angles between axes of coordinate system before and after rotation}\label{Tab1}
\fontsize{10}{14}\selectfont{
\begin{tabular}{c|c|c|c|c} \hline 
 \multicolumn{2}{c|}{\ } & \multicolumn{3}{|c}{$Axes {\ } before {\ } rotation$} \\ \cline{3-5}
\multicolumn{2}{c|}{\ } & $X$ & $Y$ & $Z$ \\ \hline \hline
$Axes$ & $X'$ & $\varphi$ & $90^0-\varphi$ & $90^0$ \\ \cline{2-5}
$after$ & $Y'$ & $90^0+\varphi$ & $\varphi$ & $90^0$ \\ \cline{2-5}
$rotation$ & $Z'$ & $90^0$ & $90^0$ & $0^0$ \\ \hline 
\end{tabular}}
\end{table}

\subsubsection{Transformation matrix on plane $\{110\}$}
\begin{figure}
\centering
\includegraphics[width=10cm]{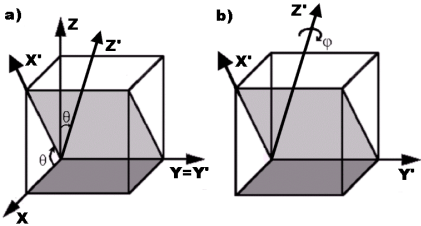}
\caption{Piezoconductance on a plane $\{110\}$: a) construction of the plane $(-101)$; b) rotation of coordinate system around $Z’$-axis.}\label{fig2}
\end{figure}

For obtaining piezoconductivity coefficients on plane $(-101)$ it is necessary to find transformation of $(001)$ to $(-101)$ plane. For this purpose, coordinate system on plane $(001)$ has to be rotated around $Y$-axis by angle $\theta=45^0$ (Fig. \ref{fig2}a). Rotation matrix has a form:

\begin{equation}\label{Eq15}
R(Y,\theta=45^0)=
\begin{bmatrix}
\cos 45^0 & 0 & \sin 45^0 \\ 
0 & 1 & 0 \\ 
-\sin 45^0 & 0 & \cos 45^0
\end{bmatrix}
\end{equation}

\noindent For investigation of piezoconductivity coefficients behavior on the plane $(-101)$, in next step it is needed further rotation of coordinate system around the new one $Z\textquotesingle$-axis by angle $\varphi$ (Fig. \ref{fig2}b). Because of silicon cell symmetry, it is enough to change $\varphi$ in a range $0\leq\varphi\leq180^0$. Matrix of this rotation has a form:

\begin{equation}\label{Eq16}
R(Z',\varphi){=}
\begin{bmatrix}
{\cos \varphi} & {\sin \varphi} & 0 \\ 
{-}{\sin \varphi\ } & {\cos \varphi\ } & 0 \\ 
0 & 0 & 1 \end{bmatrix}
\end{equation}

\noindent Resultant rotation of coordinate system can be described as a product of both mentioned above rotations (\ref{Eq15}) and (\ref{Eq16}) \cite{JFN57} \cite{Gni05}:

\begin{equation}\label{Eq17}
a=R(Z',\varphi)\cdot R(Y,45^0).
\end{equation}
\subsubsection{Transformation matrix on plane $\{111\}$}
\begin{figure}
\centering
\includegraphics[width=9cm]{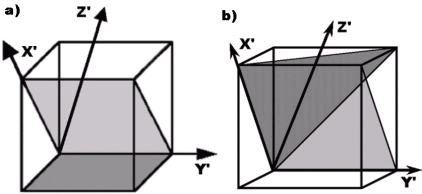}
\caption{Piezoconductance on a plane $\{111\}$: a) the plane $(-101)$; b) construction of the plane $(-111)$ via rotation of coordinate system around $X\textquotesingle$-axis.}\label{fig3}
\end{figure}

\noindent Plane $(-101)$ can be formed from plane $(001)$ via rotation around $Y$-axis by angle $\theta=45^0$ (Fig. 3a). This rotation is described by matrix (\ref{Eq15}). Plane $(-111)$ can be obtained via rotation of above-mentioned plane $(-101)$ around $X\textquotesingle$-axis by angle $\phi$ (Fig. \ref{fig3}b). From geometrical consideration, it can be concluded that $\tan{\phi}=\sqrt{2}/3$. Thus, angle $\phi$ is equal to $35.26^0$. Therefore $cos{\phi}=\sqrt{2/3}$ and $sin{\phi}=\sqrt{3}/3$ . Suitable transformation matrix has a form:

\begin{equation}\label{Eq18}
R(X',\phi){=}
\begin{bmatrix}
1 & 0 & 0 \\
0 & {\cos \phi} & {\sin \phi} \\ 
0 & {-}{\sin \phi\ } & {\cos \phi\ } 
\end{bmatrix}
\end{equation}
\noindent On Fig. \ref{fig3}b, the grey triangle is located on the new one $(-111)$ plane. Coordinate system on this plane is not marked on the picture, because of legibility. For investigation of piezoconductivity coefficients behavior on this plane, in next step it is needed further rotation of coordinate system around the new one $Z\textquotesingle\textquotesingle$-axis (not marked of Fig. \ref{fig3}) by angle $\varphi$. Because of silicon cell symmetry, it is enough to change angle $\varphi$ in range $0\leq\varphi\leq120^0$. Matrix of this rotation has a form:

\begin{equation}\label{Eq19}
R(Z'',\varphi){=}
\begin{bmatrix}
{\cos \varphi} & {\sin \varphi} & 0 \\ 
{-}{\sin \varphi\ } & {\cos \varphi\ } & 0 \\ 
0 & 0 & 1 \end{bmatrix}
\end{equation}

\noindent Resultant matrix of coordinate system rotation can be described as a product of mentioned above rotations (\ref{Eq15}) and (\ref{Eq18}) as well as (\ref{Eq19}) \cite{JFN57} \cite{Gni05}:

\begin{equation}\label{Eq20}
a=R(Z'',\varphi)\cdot R(X',\phi=35.26^0)\cdot R(Y,\theta=45^0)).
\end{equation}

\subsection{Distributions of piezoconductance coefficients}
For monocrystalline silicon, in coordinate system conformable to elementary silicon cell, piezoconductivity matrix has a form:

\begin{equation}\label{Eq21}
\Pi = -
\begin{bmatrix}
{\pi }_{11} & {\pi }_{12} & {\pi }_{12} & 0 & 0 & 0 \\
{\pi }_{12} & {\pi }_{11} & {\pi }_{12} & 0 & 0 & 0 \\
{\pi }_{12} & {\pi }_{12} & {\pi }_{11} & 0 & 0 & 0 \\ 
0 & 0 & 0 & {\pi }_{44} & 0 & 0 \\ 
0 & 0 & 0 & 0 & {\pi }_{44} & 0 \\ 
0 & 0 & 0 & 0 & 0 & {\pi }_{44}
\end{bmatrix}
\end{equation}

\begin{figure}
\centering
\includegraphics[width=10cm]{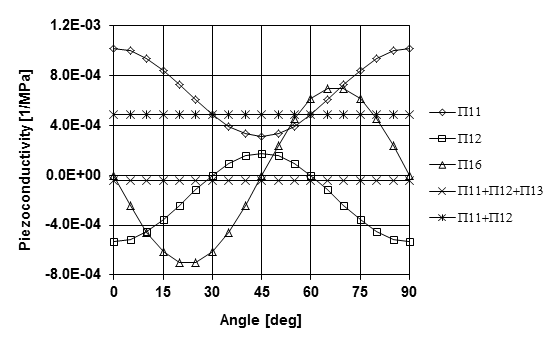}
\caption{Non-vanished piezoconductance coefficients versus angle of rotation around $Z$-axis. Plane $ (001) $, $N$-type. The angles $0^0$ and $90^0$ are equivalent to crystallographic direction $<100>$. Angle $45^0$ is equivalent to crystallographic direction $ <110>$.}\label{fig4}
\end{figure}

\begin{figure}
\centering
\includegraphics[width=10cm]{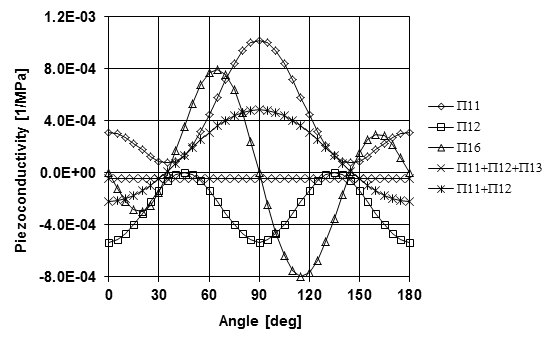}
\caption{Non-vanished piezoconductance coefficients versus angle of rotation around $Z$-axis. Plane $(-101) $, $N$-type. The angles $0^0$, $35.26^0$, $90^0$, $144.74^0$ and $180^0$ are equivalent to crystallographic directions $[101]$, $[111]$, $[010]$, $[-11-1]$ and $[-10-1]$, respectively.}\label{fig5}
\end{figure}

\begin{figure}
\centering
\includegraphics[width=10cm]{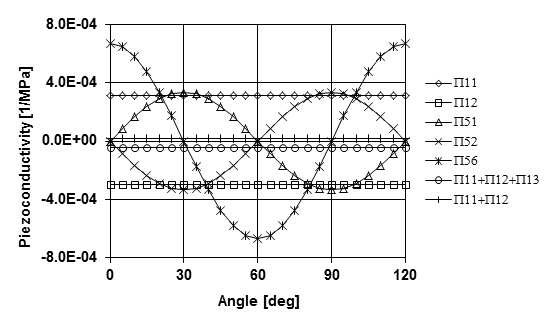}
\caption{Non-vanished piezoconductance coefficients versus angle of rotation around $Z$-axis. Plane $\{111\}$, $N$-type. The angles $0^0$, $30^0$, $60^0$, $90^0$ and $120^0$ on these plots are equivalent to crystallographic directions $[101]$, $[112]$, $[011]$, $[-121]$ and $[-110]$, respectively.}\label{fig6}
\end{figure}

\begin{figure}
\centering
\includegraphics[width=10cm]{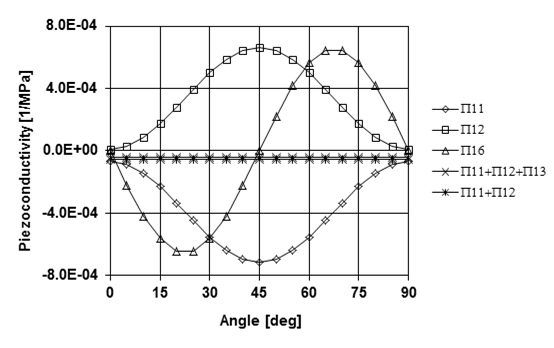}
\caption{Non-vanished piezoconductance coefficients versus angle of rotation around $Z$-axis. Plane $(001)$, $P$-type. The angles $0^0$ and $90^0$ are equivalent to crystallographic direction $<100>$. Angle $45^0$ is equivalent to crystallographic direction $<110>$.}\label{fig7}
\end{figure}

\begin{figure}
\centering
\includegraphics[width=10cm]{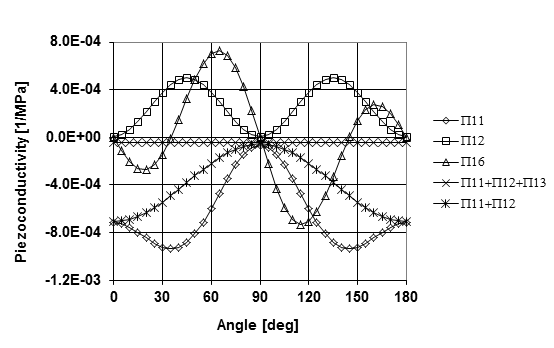}
\caption{Non-vanished piezoconductance coefficients versus angle of rotation around $Z$-axis. Plane $(-101)$, $P$-type. The angles $0^0$, $35.26^0$, $90^0$, $144.74^0$ and $180^0$ are equivalent to crystallographic directions $[101]$, $[111]$, $[010]$, $[-11-1]$ and $[-10-1]$, respectively.}\label{fig8}
\end{figure}

\begin{figure}
\centering
\includegraphics[width=10cm]{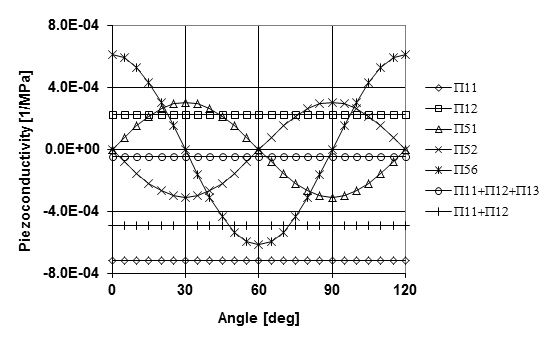}
\caption{Non-vanished piezoconductance coefficients versus angle of rotation around $Z$-axis. Plane $(-1-11)$, $P$-type. The angles $0^0$, $30^0$, $60^0$, $90^0$ and $120^0$ on these plots are equivalent to crystallographic directions $[101]$, $[112]$, $[011]$, $[-121]$ and $[-110]$, respectively.}\label{fig9}
\end{figure}

\noindent In engineering practice, the values of $\pi_{11}$, $\pi_{12}$ and $\pi_{44}$ should be extracted from the measurements, for bulk or inversion layers respectively. Theirs values for different types of conducting layers are given in literature \cite{CSS54} \cite{Can791} \cite{Col682} \cite{Mik811}. Using data from a variety of sources, different piezoconductivity coefficients will be obtained. Without loss of generality, coefficients $\pi_{11}$, $\pi_{12}$ and $\pi_{44}$ given by Smith \cite{CSS54} will be used for farther analysis (Table \ref{Tab2}). On Figures \ref{fig4}--\ref{fig9}, non-vanished piezoconductivity coefficients and specific sums on planes $\{100\}$, $\{110\}$ and $\{111\}$ are presented both for holes and for electrons. Analysis of distributions of piezoconductivity coefficients shows how to locate the MOSFET channel.

\begin{table}
\centering
\caption{Piezoresistance coefficients for silicon \cite{CSS54}.}\label{Tab2}
\fontsize{10}{14}\selectfont{
\begin{tabular}{c|c|c|c} \hline 
 & $\pi_{11} [MPa^{-1}]$ & $\pi_{12} [MPa^{-1}]$ & $\pi_{44} [MPa^{-1}]$ \\ \hline \hline 
P-Si & 66 & -11 & 1381 \\ \hline 
N-Si & -1022 & 534 & -136 \\ \hline 
\end{tabular}}
\end{table}

\subsubsection{The most promising MOSFET location}
From analysis of piezoconductance distributions (Figs. \ref{fig4}-\ref{fig9}) the most promising channel locations on different planes will be selected, for assumed uniaxial stress state:
\begin{enumerate}
\item $N$-channel on plane $\{100\}$: for direction $<100>$ ($0^0$ or $90^0$ on Fig. \ref{fig4}) maximum increment of conductivity can be obtained with positive uniaxial stress along channel – because of positive value of $\Pi_{11}$ maximum.

\item $N$-channel on plane $\{110\}$: for direction $<100>$ ($90^0$ on Fig. \ref{fig5}) desirable increment of conductivity can be obtained under uniaxial longitudinal positive stress, because of positive $\Pi_{11}$ maximum.

\item $N$-channel on plane $\{111\}$: for direction $<112>$ ($30^0$ on Fig. \ref{fig6}) occurs beneficial superposition of $\kappa_1$ and $\kappa_5$. Maximum increment of conductivity can be achieved under longitudinal positive (see values of $\Pi_{11}$ and $\Pi_{51}$) or transversal negative (see values of $\Pi_{12}$ and $\Pi_{52}$) stress.

\item $P$-channel on planes $\{100\}$: maximum increment of conductivity for $<110>$ direction ($45^0$ on Fig. \ref{fig7}) could be obtained for two cases of stress state:
\begin{enumerate}
\item For uniaxial negative (compressed) stress along $X$-axis (along channel) – because of negative minimum of $\Pi_{11}$.
\item For uniaxial positive (tensile) stress along $Y$-axis (in plane of channel, perpendicularly to channel length) – because of positive maximum of $\Pi_{12}$.
\end{enumerate}

\item $P$-channel on planes $\{110\}$: maximal increment of conductivity could be performed for $<111>$ direction ($35.26^0$ or $144.74^0$ on Fig. \ref{fig8}), under uniaxial longitudinal negative stress, because of $\Pi_{11}$ negative minimum.

\item P-channel on planes $\{111\}$: for $<112>$ direction ($90^0$ on Fig. \ref{fig9}) occurs beneficial superposition of $\kappa_1$ and $\kappa_5$. Drain current can be raised under longitudinal negative stress (coefficient $\Pi_{11}$ and $\Pi_{51}$) or under transversal positive stress (coefficient $\Pi_{12}$ and $\Pi_{52}$). The first one event is more profitable because of $|\Pi_{11}|>|\Pi_{12}|$.

\end{enumerate}
\noindent Summary results of these analyzes are presented in Table \ref{Tab3}.

\begin{table}
\centering
\caption{The most promising MOSTET location for the uniaxial state of stress.}\label{Tab3}
\fontsize{10}{14}\selectfont{
\begin{tabular}{c|c|c|c|c|c} \hline 
Channel & Channel & \multirow{2}{*}{Angle} & \multirow{2}{*}{Reference} & Nonzero & \multirow{2}{*}{Cause} \\ 
 type & location & & & stress component & \\ \hline \hline
\multirow{4}{*}{N} & $[100]/(001)$ & $0^0$ or $90^0$ & Fig. \ref{fig4} & $\sigma_1>0$ & $\max\Pi_{11}$ \\ \cline{2-6}
 & $[010]/(-101)$ & $90^0$ & Fig. \ref{fig5} & $\sigma_1>0$ & $\max\Pi_{11}$ \\ \cline{2-6}
 & \multirow{2}{*}{$[112]/(-111)$} & \multirow{2}{*}{$30^0$} & \multirow{2}{*}{Fig. \ref{fig6}} & $\sigma_1>0$ & $\max\Pi_{51}$, $\Pi_{11}>0$ \\ \cline{5-6}
 & & & & $\sigma_2<0$ & $\min\Pi_{52}$, $\Pi_{12}<0$ \\ \hline
\multirow{6}{*}{P} & \multirow{2}{*}{$[110]/(001)$} & \multirow{2}{*}{$45^0$} & \multirow{2}{*}{Fig. \ref{fig7}} & $\sigma_1<0$ & $\min{\Pi_{11}}$ \\ \cline{5-6}
 & & & & $\sigma_2>0$ & $\max\Pi_{12}$ \\ \cline{2-6}
 & [111]/(-101) & $35.26^0$ & \multirow{2}{*}{Fig. \ref{fig8}} & \multirow{2}{*}{$\sigma_1<0$} & \multirow{2}{*}{$\min\Pi_{11}$} \\ \cline{2-3}
 & [-11-1]/(-101) & $144.74^0$ & & & \\ \cline{2-6}
 & \multirow{2}{*}{$[-121]/(-111)$} & \multirow{2}{*}{$90^0$} & \multirow{2}{*}{Fig. \ref{fig9}} & $\sigma_1<0$ & $\min\Pi_{51}$, $\Pi_{11}<0$ \\ \cline{5-6}
 & & & & $\sigma_2>0$ & $\max\Pi_{52}$, $\Pi_{12}>0$ \\ \hline
\end{tabular}}
\end{table}

\section{General model of stressed MOSFET}
Switched on MOSFET is investigated. MOSFET channel is located on arbitrary chosen crystallographic plane. Cartesian coordinate system associated with the MOSFET will be further considered. Transistor channel can be treated as a conducting layer located along $X$-axis of this coordinate system. $Y$-axis is located horizontally in a plane of channel. It is transversal to the channel length. $Z$-axis is vertical to the plane created by length and width of the MOS transistor (Fig. 10). In the channel, there exists a vector of electric field $E$ and density current $j$, because of transistor polarization. Such a coordinate system gives adequate signs of specific components of electric field and current density. In normal polarization, in $n$-type MOSFET this components are positive and in $p$-type – negative.

\begin{figure}
\centering
\includegraphics[width=10cm]{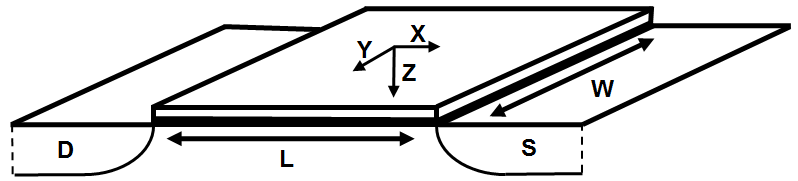}
\caption{Schematic diagram of MOSFET.}\label{fig10}
\end{figure}

Component $E_1$ of electric field is parallel to $X$-axis. Vertical component $E_3$ is parallel to $Z$-axis. In plane of channel, component $E_2$ parallel to $Y$-axis can be also respected. In MOS transistor channel, density current $j_1$ parallel to $X$-axis is considered. This component is dependent on all nonzero components of electric field. Based on formulas (\ref{Eq1}) and (\ref{Eq2}) it can be expressed:

\begin{equation}\label{Eq22}
j_1=\kappa_1E_1+\kappa_6E_2+\kappa_5E_3.
\end{equation}

\noindent Only three components $\kappa_1$, $\kappa_5$ and $\kappa_6$ of conductivity (\ref{Eq2}) are inherent in this equation. From (\ref{Eq4}), (\ref{Eq6}) and (\ref{Eq7}), these components can be expressed as follows:

\begin{equation}\label{Eq23}
\begin{cases}
\kappa_1=\kappa+\Delta\kappa_1 \\
\kappa_5=\Delta\kappa_5 \\
\kappa_6=\Delta\kappa_6
\end{cases}
\end{equation}

\noindent Using (\ref{Eq23}), density current (\ref{Eq22}) can be articulated as:
\begin{equation}\label{Eq24}
j_1=\kappa{E_1}+\Delta\kappa_1E_1+\Delta\kappa_6E_2+\Delta\kappa_5E_3.
\end{equation}
\noindent Whole drain current can be obtained from integration of density component $j_1$ over transistor channel cross-section $S$ \cite{Sah05}:

\begin{equation}\label{Eq25}
I_D=\int\limits_{S}j_1dS=\int\limits_{S}(\kappa{E_1}+\Delta\kappa_1E_1+\Delta\kappa_6E_2+\Delta\kappa_5E_3)dS.
\end{equation}

\noindent It can be noticed that $\kappa$ is an effective conductivity of unstressed MOSFET. On the other hand, relative change of conductivity is dependent on stress in MOSFET channel. According to (\ref{Eq8}), changes of conductivity components $\Delta\kappa_1$, $\Delta\kappa_5$ and $\Delta\kappa_6$ can be expressed as follows:

\begin{equation}\label{Eq26}
\begin{cases}
\Delta\kappa_1=\kappa\sum\limits_{j=1}^6\Pi_{1j}\sigma_j \\
\Delta\kappa_5=\kappa\sum\limits_{j=1}^6\Pi_{5j}\sigma_j \\
\Delta\kappa_6=\kappa\sum\limits_{j=1}^6\Pi_{6j}\sigma_j
\end{cases}
\end{equation}

Substituting (\ref{Eq26}) to (\ref{Eq25}), general model of drain current has a form:

\begin{equation}\label{Eq27}
I_D=\int\limits_{S}{\left(
\kappa{E_1}
+\kappa E_{1}\sum^6_{j=1}{\Pi_{1j}\sigma_j}
+\kappa E_{2}\sum^6_{j=1}{\Pi_{6j}\kappa_j}
+\kappa E_{3}\sum^6_{j=1}{\Pi_{5j}\kappa_j}
\right)dS}.
\end{equation}

\noindent If no stress is assumed, then components $\Delta\kappa_1$, $\Delta\kappa_5$ and $\Delta\kappa_6$ are vanishing and model (\ref{Eq27}) has no anisotropic components. It describes (denoted as $I_{D0}$) MOSFET drain current without stress in the channel:

\begin{equation}\label{Eq28}
I_{D0}=\int\limits_{S}{j_1dS}=\int\limits_{S}{\kappa{E_1}dS}.
\end{equation}

\noindent All this means that drain current (\ref{Eq27}) can be described as a superposition of two components. The first one $I_{D0}$ is not dependent on stress. The second one denoted as ${\Delta}I_D$ is an anisotropic component dependent on stress:

\begin{equation}\label{Eq29}
{\Delta}I_D=\int\limits_{S}{\left(
\kappa E_{1}\sum^6_{j=1}{\Pi_{1j}\sigma_j}
+\kappa E_{2}\sum^6_{j=1}{\Pi_{6j}\sigma_j}
+\kappa E_{3}\sum^6_{j=1}{\Pi_{5j}\sigma_j}
\right)dS}.
\end{equation}

\noindent This way, whole drain current can be expressed as a sum of these two components:
\begin{equation}\label{Eq30}
I_D=I_{D0}+{\Delta}I_D.
\end{equation}

For modeling drain current without stress, only $I_{D0}$ can be used. In practical application, this component can be substituted by any of numerical or analytical correct model, if necessary. Model of $I_{D0}$ should take into consideration all complicated physical aspects inherent in MOSFET: saturation region, short channel, mobility vs. velocity, etc. For instance, in analytical mode it can be modeled with the use of any SPICE like model. Component ${\Delta}I_D$ allow incorporating of stress, in any of $I_{D0}$ model without stress effect.

For increment of current efficiency, in formula (\ref{Eq29}) positive increment of conductivity (\ref{Eq26}) is necessary. This increment will be positive if signs of stress components will be the same like signs of specific piezoconductivity coefficients.

\section{Model simplification}
Specific properties of considered structure allows simplify component (\ref{Eq29}). Simplifications resulted from stress state in transistor channel, MOS structure symmetry as well as analytical approach will be taken into consideration in further analysis.

\subsection{Specific stress state in MOSFET channel}
In formula (\ref{Eq29}), there are components of piezoconductance coefficients $\Pi_{ij}$ with first index $i$ equal to $1$, $5$ or $6$. Selection of second index $j$ is determined by nonzero $\sigma_j$ stress components inherent in the channel. In general case of stress state, indexes $j$ can have all possible values from $1$ to $6$. For further analysis, some specific cases of stress state in plane of MOSFET channel \cite{Mai071}, against a general case of stress state will be considered:

\begin{itemize}
\item Isotropic stress induced by hydrostatic pressure $p$ $–-$ indexes $j$ are equal to $1$, $2$ and $3$.
\item Isotropic stress in a plane of channel $–-$ indexes $j$ are equal to $1$ and $2$.
\item Uniaxial stress along channel $–-$ index $j$ is equal to $1$.
\item Uniaxial stress across channel $–-$ index $j$ is equal to $2$.
\item General case of flat stress in a plane of channel $–-$ indexes $j$ are equal to $1$, $2$ and $6$.
\end{itemize}

\noindent Changes of conductivity component $\Delta\kappa_i$ versus specific stress state are presented in Table \ref{Tab4}. From here, it can be concluded which of piezoconductance coefficients are indispensable for modeling of stress influence on MOSFET drain current. These coefficients should be taken into consideration in MOS transistor drain current modeling.

\begin{table}
\centering
\caption{Change of i-th conductivity component as a function of stress state.}\label{Tab4}
\fontsize{10}{14}\selectfont{
\begin{tabular}{c|c|c} \hline 
Stress state & Stress in vector notation & Change of conductivity \\ \hline \hline
General case & $\sigma=\left[\sigma_1,\sigma_2,\sigma_3,\sigma_4,\sigma_5,\sigma_6\right]^T $ & $\Delta\kappa_i=\kappa\sum\limits_{j=1}^{6}{\Pi_{ij}\sigma_j} $ \\ \hline
Hydrostatic & \multirow{2}{*}{$\sigma=\left[-p,-p,-p,0,0,0\right]^T $} & \multirow{2}{*}{$\Delta\kappa_i=-\kappa{p} \sum\limits_{j=1}^{3}{\Pi_{ij}} $} \\ 
 pressure $p$ & & \\ \hline
Isotropic stress & \multirow{2}{*}{$\sigma=\left[\sigma_0,\sigma_0,0,0,0,0\right]^T $} & \multirow{2}{*}{$\Delta\kappa_i=\kappa\sigma_0 \left(\Pi_{i1}+\Pi_{i2} \right) $} \\ 
in plane of channel & & \\ \hline
 Uniaxial stress & \multirow{2}{*}{$\sigma=\left[\sigma_1,0,0,0,0,0\right]^T$} & \multirow{2}{*}{$ \Delta\kappa_i=\kappa\sigma_1\Pi_{i1}$} \\
 along channel & & \\ \hline
Uniaxial stress & \multirow{2}{*}{$\sigma=\left[0,\sigma_2,0,0,0,0\right]^T $} & \multirow{2}{*}{$\Delta\kappa_i=\kappa\sigma_2\Pi_{i2}$} \\ 
across channel & & \\ \hline
General case & \multirow{3}{*}{$\sigma=\left[\sigma_1,\sigma_2,0,0,0,\sigma_6\right]^T $} & \multirow{3}{*}{$\Delta\kappa_i=\kappa\left(\Pi_{i1}\sigma_1+\Pi_{i2}\sigma_2+\Pi_{i6}\sigma_6\right) $} \\
of flat stress & & \\
in plane of channel & & \\ \hline
\end{tabular}}
\end{table}

\subsection{Symmetry of MOSFET structure}
Classical MOSFET is symmetric relatively to $XZ$ plane. Assumption about symmetry of MOS transistor structure implies supposition about symmetry of distributions of theirs parameters. In particular, distributions of piezoconductance coefficients and stress components are also symmetric. It means that these quantities are even functions of variable $Y$. On the other hand, component $E_2$ of electric field along $Y$-axis has opposite signs near opposite edges and it is vanished in the middle of the structure. It can be supposed that it is an odd function of variable $Y$ with average value equal to zero. If such assumptions are reliable then suitable component in (\ref{Eq27}) can be vanished:

\begin{equation}\label{Eq31}
\int\limits_{S}{\left(
\kappa E_{2}\sum^6_{j=1}{\Pi_{6j}\sigma_j}
\right)dS}=0.
\end{equation}

\noindent This way model (\ref{Eq27}) can be simplified to the form without piezoconductance coefficients with first index equal to $6$:

\begin{equation}\label{Eq32}
I_D=I_{D0}+\Delta{I_D}=
\int\limits_{S}{\left(
\kappa E_1
+\kappa E_{1}\sum^6_{j=1}{\Pi_{1j}\sigma_j}
+\kappa E_{3}\sum^6_{j=1}{\Pi_{5j}\sigma_j}
\right)dS}.
\end{equation}

\noindent Three-dimensional ($3-D$) model of $\Delta{I_D} $ vs. electric field is reduced to a two-dimensional ($2-D$) problem. This result is conformable with principle of $2-D$ compact model \cite{Sah05}.

\subsection{Analytical approach}
Model (\ref{Eq32}) gives possibility to take into consideration the influence of the stress to the MOSFET current. This model has a general capacity. On the other hand, analytical compact model is necessary in circuit simulator \cite{Sah05}. Special problem is adequate transformation of (\ref{Eq32}) to analytical form. Analytical model contains semi-empirical and empirical parameters, which must be extracted. In general, it contains two types of parameters: process parameters and electrical parameters. Process parameters must be taken from process characterization. Electrical parameters must be determined from measured electrical data \cite{Fot99}. Additionally, model that takes into consideration influence of stress should contain information about stress and piezoconductivity coefficients. In general, both of these parameters can be non-homogeneous in the channel. In compact analytical modeling, non-uniform distributions of parameters must be substituted by representative constant values. For modeling of bulk piezoresistive devices average stress and extracted effective piezoresistivity (piezoconductivity) coefficients are used \cite{Gni06}. The same procedure can be employed for this paper approximation. If real stress and piezoconductance coefficients distributions are substituted by theirs effective values (it means values that are not a function of integrated area), then suitable sums of products $\Pi_{ij}\sigma_j$ can be drawing out before integrals.

Similarly, if $\kappa$ is an effective conductivity of unstressed MOSFET then it can be drawing out before integrals too. This way, integrals of electric field components over cross-section $S$ can be substituted by products of $S$ and average values of electric field components in mentioned cross-section. From here, equation (\ref{Eq32}) can be still simplified to compact analytical form:

\begin{equation}\label{Eq33}
I_D=I_{D0}{\left(
1+\sum^6_{j=1}{\Pi_{1j}\sigma_j}
\right)}
+S\kappa{E_3}\sum^6_{j=1}{\Pi_{5j}\sigma_j}.
\end{equation}
\subsection{Models of drain current on selected planes}
For assumed symmetry of MOSFET and assumed planes of structure location as well as stress distribution, model (\ref{Eq33}) can be further simplified. For this purpose, we assume that MOSFET channel can be located on typical planes: $\{100\}$, $\{110\}$ and $\{111\}$. For that reason, distribution of piezoconductance coefficients should be considered on these planes. In our approach, only piezoconductance coefficients with first indexes $1$ and $5$ will be examined, because of assumed structure symmetry. Second indexes are determined by assumptions about stress state in the channel.

From Table \ref{Tab4} it follows, that for assumed flat stress state in the channel, only coefficients with second indexes equal to $1$, $2$ and $6$ are important. Moreover, sums $(\Pi_{11}+\Pi_{12}+\Pi_{13})$ and $(\Pi_{51}+\Pi_{52}+\Pi_{53})$ have to be investigated, for hydrostatic pressure. Additionally, sums $(\Pi_{11}+\Pi_{12})$ and $(\Pi_{51}+\Pi_{52})$ are significant for isotropic stress state in a plane of channel. Analysis of distribution of specific coefficients on different planes can show which directions of transistor channel location are desirable for improvement of MOSFET performances. For assumed specific stress state model of drain current (\ref{Eq33}) can be simplified to the forms presented in Table \ref{Tab5}. 

\begin{table}
\centering
\caption{Analytical models of MOSFET current.}\label{Tab5}
\fontsize{10}{14}\selectfont{
\begin{tabular}{c|c|c|c} \hline 
 \multirow{2}{*}{No.} & \multirow{2}{*}{State of stress} & \multirow{2}{*}{Form of equation (\ref{Eq33})} & Plane \\ 
 & & & of channel \\ \hline \hline
 \multirow{2}{*}{(M1)} & Hydrostatic & \multirow{2}{*}{$I_D=I_{D0}\left(1-p\left(\Pi_{11}+\Pi_{12}+\Pi_{13}\right)\right)$} &
 \multirow{2}{*}{All} \\ 
 & pressure $p$ & & \\ \hline
\multirow{2}{*}{(M2)} & Isotropic stress & $I_D=I_{D0} \left(1+\sigma_0\left(\Pi_{11}+\Pi_{12} \right)\right)$ & $\{100\}, \{110\}$ \\ \cline{3-4}
 & in plane of channel & $I_D=I_{D0} \left(1+\sigma_0\left(\Pi_{11}+\Pi_{12}+\Pi_{13}\right)\right)$ & $\{111\}$ \\ \hline
 \multirow{2}{*}{(M3)} & Uniaxial stress & $I=I_{D0}\left(1+\Pi_{11}\sigma_1\right) $ & $\{100\}, \{110\}$ \\\cline{3-4}
 & along channel & $I=I_{D0}\left(1+\Pi_{11}\sigma_1\right)+S\kappa{E_3}\Pi_{51}\sigma_1 $ & $\{111\}$ \\ \hline
\multirow{2}{*}{(M4)} 
& Uniaxial stress & $I=I_{D0} \left(1+\Pi_{12}\sigma_2 \right) $ & $\{100\}, \{110\}$ \\ \cline{3-4}
& across channel & $I=I_{D0} \left(1+\Pi_{12}\sigma_2 \right)+S\kappa{E_3}\Pi_{52}\sigma_2 $ & $\{111\}$ \\ \hline
\multirow{2}{*}{(M5)}
 & Flat stress state & 
 $I=I_{D0}\left(1+\Pi_{11}\sigma_1+\Pi_{12}\sigma_2+\Pi_{16}\sigma_6 \right) $ & $\{100\}, \{110\}$ \\ \cline{3-4}
 & in plane of channel & 
 $I=I_{D0}\left(1+\Pi_{11}\sigma_1+\Pi_{12}\sigma_2+\Pi_{16}\sigma_6 \right) $ & \multirow{2}{*}{$\{111\}$}\\
 & $-$general case & 
 $+S\kappa{E_3}\left(\Pi_{51}\sigma_1+\Pi_{52}\sigma_2+\Pi_{56}\sigma_6 \right) $ & \\\hline
\end{tabular}
}
\end{table}
\section{Examples}
As an example, analytical approach will be presented based on the formulas $(M2)$ and $(M3)$ in Table \ref{Tab5}. For this purpose, uniaxial stress with a value of $200MPa$ will be considered in transistor channel. For calculation of drain current under stress, it is necessary to know specific piezoconductance coefficients and drain current $I_{D0}$. Additionally, on plane $\{111\}$ effective conductivity κ in MOSFET channel as well as vertical component of electric field $E_3$ should be known. Conductivity $\kappa$ will be calculated as normalized output conductance $G_{DS}$:

\begin{equation}\label{Eq34}
\kappa=G_{DS}\cdot\frac{L}{S}=\frac{\partial{I_{D0}}}{\partial{V_{DS}}}\cdot\frac{L}{S}.
\end{equation}

In plane $\{111\}$ two types of the stress impact on the MOSFET drain current are observed. The first type is associated with the impact of stress on the conductivity component $\kappa_1$, through coefficients $\Pi_{11}$ or $\Pi_{12}$ (respectively for longitudinal or transverse stress). The second one is associated with the impact of the stress on the conductivity component $\kappa_5$, through coefficients $\Pi_{51}$ or $\Pi_{52}$ (respectively for longitudinal or transverse stress). Conductivity component $\kappa_1$ combines current density $j_1$ with electric field component $E_1$. Conductivity component $\kappa_5$ combines current density $j_1$ with vertical electric field component $E_3$. It would be interesting to examine the nature and scope of these two interactions.

\subsection{N-channel MOSFET under stress}
\begin{figure}
\centering
\includegraphics[width=10cm]{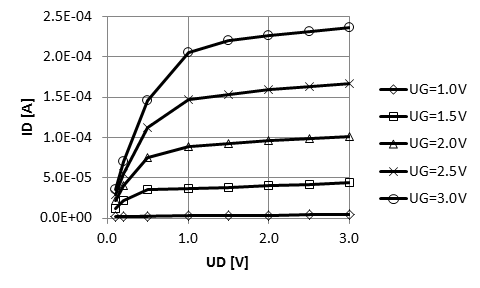}
\caption{N-MOSFET drain current without stress.}\label{fig11}
\end{figure}

For calculation influence of stress, classical n-channel MOSFET was modeled in $2-D$ mode, with the use of Minimos $6.1$ Win simulator \cite{Minimos}. For calculation it was assumed oxide thickness $T_{OX}=15nm$, channel width $W=1\mu{m} $, channel length $L=0.85\mu{m} $. Calculated characteristics are shown on Fig. \ref{fig11}.

\begin{figure}
\centering
\includegraphics[width=10cm]{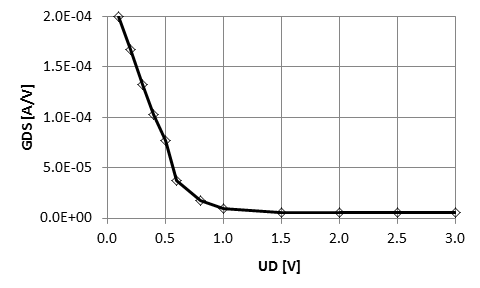}
\caption{Conductance $G_{DS}$ under $U_G=2V$.}\label{fig12}
\end{figure}

For proposed methodology, stress influence under $UG=2V$ was considered. Results of calculation was used for approximation of effective values of conductivity, vertical component of electric field $E_3$ and drain current $I_{D0}$. Thickness of inversion layer was assumed $x_d=0.15\mu{m} $. Area of transistor cross-section was approximated as $S=x_d\cdot{W}=0.15\mu{m} ^2$. Vertical electric field was assumed $E_z=0.23V/\mu{m} $. Calculated conductance $G_{DS}$ under $U_G=2V$ is presented on Fig. \ref{fig12}.

From analysis of piezoconductance distributions (Figs. \ref{fig4}, \ref{fig6} and \ref{fig8}) for $N$-channel MOSFET, on specific planes $\{100\}$, $\{110\}$ and $\{111\}$ the most promising channel locations are along $<100>$, $<100>$ and $<121>$ directions respectively (see Table \ref{Tab3}). With these assumptions, the drain current under stress was calculated and presented in Figure \ref{Eq13}. The best results are obtained for channel located on $(001)$ along $[100]$ under tensile longitudinal stress. The same results are obtained for channel location on $(-101)$ plane along $[010]$ direction, also for tensile longitudinal stress. The results on the plane $\{111\}$, for both of tensile longitudinal stress and compressive transversal stress are nearly identical, with a slight predominance of the former.

\begin{figure}
\centering
\includegraphics[width=10cm]{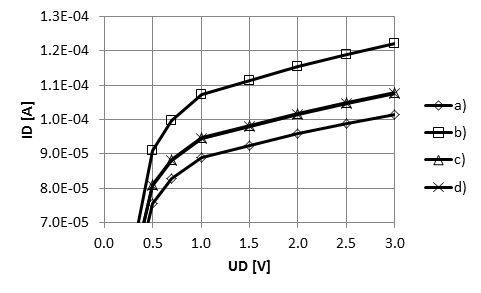}
\caption{Influence of $200 MPa$ uniaxial stress on $N$-MOSFET characteristics: a) $U_G=2V$, no stress; b) channel $[100]/(001)$ or $[010]/(-101)$, tensile longitudinal stress; c) channel $[112]/(-111)$, tensile longitudinal stress; d) channel $[112]/(-111)$, compressive transversal stress.}\label{fig13}
\end{figure}

\begin{figure}
\centering
\includegraphics[width=10cm]{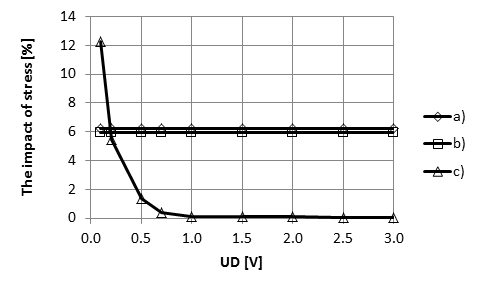}
\caption{The impact of the stress on the growth of current for $N$-MOSFET in channel $[112]/(-111)$: a) impact of tensile longitudinal stress via conductance $\kappa_1$; b) impact of compressive transversal stress via conductance $\kappa_1$; c) impact of tensile longitudinal stress via conductance $\kappa_5$ or impact of compressive transversal stress via conductance $\kappa_5$.}\label{fig14}
\end{figure}

N-MOSFET current change associated with the change of the conductivity components under stress were examined on plane $\{111\}$ (Fig. \ref{fig14}). It was found that the change in conductivity $\kappa_1$ increases the MOSFET current by about six percent. Meanwhile, the change of MOSFET current caused by a change of conductivity $\kappa_5$ is non-linear. In the linear part of the MOSFET characteristics, change of current is more than twelve percent. When the MOSFET current comes into saturation region, these changes are disappearing.

\subsection{P-channel MOSFET under stress - thought experiment}
To compare the effects of stress on the characteristics of $p$-channel MOSFET, a thought experiment will be tested, assuming the same current efficiency as previously analyzed $N$-MOSFET. In addition, it is assumed $U_G=-2V$, area of transistor cross-section $S=0.15\mu{m} ^2$ and vertical electric field $E_z=0.23V/\mu{m} $.

\begin{figure}
\centering
\includegraphics[width=10cm]{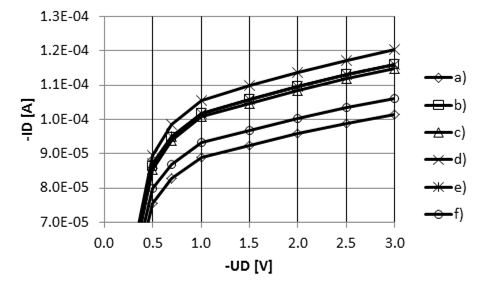}
\caption{Influence of 200 MPa uniaxial stress on P-MOSFET characteristics: a) UG=2V, no stress; b) channel [110]/(001), compressive longitudinal stress; c) channel [110]/(001), tensile transversal stress; d) channel [111]/(-101) or [-11-1]/(-101), tensile longitudinal stress; e) channel [-121]/(-111), compressive longitudinal stress; f) channel [-121]/(-111), tensile transversal stress.}\label{fig15}
\end{figure}

From analysis of piezoconductance distributions for p-channel MOSFET (Figs. \ref{fig5}, \ref{fig7} and \ref{fig9}), on specific planes $\{100\}$, $\{110\}$ and $\{111\}$ the most promising channel locations are along $<110>$, $<111>$ and $<121>$ directions, respectively.

\begin{figure}
\centering
\includegraphics[width=10cm]{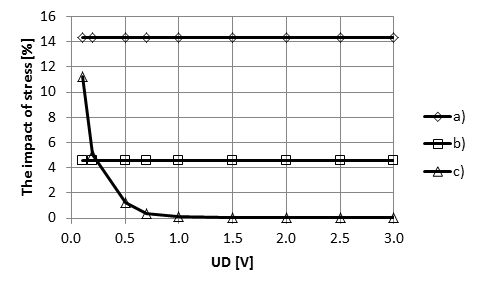}
\caption{The impact of the stress on the growth of current for $P$-MOSFET in channel $[-112]/(-111)$: a) impact of compressive longitudinal stress via conductance $\kappa_1$; b) impact of tensile transversal stress via conductance $\kappa_1$; c) impact of tensile transversal stress via conductance $\kappa_5$ or impact of compressive longitudinal stress via conductance $\kappa_5$.}\label{fig16}
\end{figure}

Obtained results are presented in Fig. \ref{fig15}. The best results are achieved for channel located on $\{110\}$ along $<111>$, under tensile longitudinal stress. On plane $\{100\}$, results obtained for compressive longitudinal stress are slightly better than results for tensile transversal stress. On plane $\{111\}$, results achieved for compressive longitudinal stress are better than results for tensile transversal stress.

The change of P-MOSFET current, associated with the change of the conductivity components under stress, were examined on plane $\{111\}$ (Fig. \ref{fig16}). The change in conductivity $\kappa_1$ under compressive longitudinal stress increases the MOSFET current by about $14\%$. On the other hand, the change in conductivity $\kappa_1$ under compressive transversal stress increases the current by about $4\%$. The change of MOSFET current caused by a change of conductivity $\kappa_5$ is non-linear. In the linear part of the MOSFET characteristics, change of current is more than ten percent. In saturation region, these changes are disappearing.

\section{Discussion}
In this paper, behavioral description of MOSFET drain current, under low mechanical load, from linear piezoconductivity point of view was considered. The approach proposed above, implies several other problems that should be more precisely explained.

\subsection{Symmetry reduction}
The proposed method does not take into account the reduction of symmetry \cite{SymRed2010}. It was assumed that description of piezoconductivity for bulk layers is also valid for inversion layer in MOSFET channel. Under this assumption, for monocrystalline bulk silicon, in coordinate system conformable to elementary silicon cell, matrix of piezoconductivity has a form:
\begin{equation}\label{Eq35}
\Pi_B =
\begin{bmatrix}
{\Pi }_{11} & {\Pi }_{12} & {\Pi }_{12} & 0 & 0 & 0 \\
{\Pi }_{12} & {\Pi }_{11} & {\Pi }_{12} & 0 & 0 & 0 \\
{\Pi }_{12} & {\Pi }_{12} & {\Pi }_{11} & 0 & 0 & 0 \\ 
0 & 0 & 0 & {\Pi }_{44} & 0 & 0 \\ 
0 & 0 & 0 & 0 & {\Pi }_{44} & 0 \\ 
0 & 0 & 0 & 0 & 0 & {\Pi }_{44}
\end{bmatrix}
\end{equation}
On the other hand in $<100>/(001)$ MOSFET, because of symmetry reduction, the number of independent piezoconductance coefficients increases to six. In such case, in coordinate system conformable to elementary silicon cell, piezoconductivity matrix takes the form \cite{SymRed2010}:

\begin{equation}\label{Eq36}
\Pi =
\begin{bmatrix}
{\Pi }_{11} & {\Pi }_{12} & {\Pi }_{13} & 0 & 0 & 0 \\
{\Pi }_{12} & {\Pi }_{11} & {\Pi }_{13} & 0 & 0 & 0 \\
{\Pi }_{13} & {\Pi }_{13} & {\Pi }_{33} & 0 & 0 & 0 \\ 
0 & 0 & 0 & {\Pi }_{44} & 0 & 0 \\ 
0 & 0 & 0 & 0 & {\Pi }_{44} & 0 \\ 
0 & 0 & 0 & 0 & 0 & {\Pi }_{66}
\end{bmatrix}
\end{equation}
Reduction of symmetry introduces disturbance in the description of the piezoconductivity. It means that piezoconductivity in MOSFET can be described as a superposition of the matter tensor and the field tensor. For that reason, both effects can be considered independently. Matrix (2) can be expressed as the sum of two matrices:
\begin{equation}\label{Eq37}
\Pi=\Pi_B+\Delta\Pi
\end{equation}
The first one is the matrix (\ref{Eq35}) that represents tensor of matter in coordinate system of elementary silicon cell. The second one is a matrix, which represents disturbance associated with reduction of the symmetry. It may be interpreted as a field tensor. The change of coordinate system changes tensor components. It is assumed, that optimal direction of the transistor channel is searched on the plane $(001)$. The matrix (\ref{Eq36}) should be rotated around the $Z$-axis, in accordance with equation (\ref{Eq13}):
\begin{equation}\label{Eq38}
\Pi^{\textquotesingle}=\alpha\Pi\alpha^{-1}=\alpha\left(\Pi_B+\Delta\Pi\right)\alpha^{-1}=\alpha\Pi_{B}\alpha^{-1}+\alpha\Delta\Pi\alpha^{-1}.
\end{equation}
There is a problem, how to take into account the reduction of symmetry on other planes? To solve this problem, first it is necessary to find matter tensor $\Pi_B$ on the given plane. For the plane $\{110\}$, the formula (\ref{Eq15}) should be used. Similarly, formulas (\ref{Eq18}) and (\ref{Eq19}) should be used, for the plane $\{111\}$. In turn, for finding the optimal direction of transistor channel, for the given field tensor $\Delta\Pi$ (in the same coordinate system), formula (\ref{Eq14}) should be used.

Because the piezoconductivity in MOSFET can be described as a superposition of material tensor and field tensor, both effects can be considered independently. Therefore, the proposed method can easily be expanded into a form that will take into account the reduction of symmetry.

\subsection{Other problems}
The approach proposed above, implies several other problems that should be more precisely explained.

First problem is a model simplification. General model (\ref{Eq27}) can be simplified for practical use.
First simplifications results from reduction of some stress components in the channel or from vanishing of a certain piezoconductance components in specific crystallographic planes.
Next simplification results from omission of electric field component $E_2$, because of symmetry of MOSFET structure assumption. If this component is omitted, then conductivity component $\kappa_{12}$ (denoted as $\kappa_6$ in vector notation) and piezoconductance coefficients with first index equal to six are omitted too. This way, $3-D$ model of $I_D$ is reduced to a $2-D$ model, in electric field space.
Last simplification is based on substitution of real distributions of stress, piezoconductance coefficients, conductivity and electric field by theirs effective values, which should be extracted (piezoconductivity coefficients) or calculated (stress components, specific electric field component, conductivity $\kappa$). These gives an analytical approach and it is desirable from point of view of circuit simulator model.
On the other hand, it can be stated that these simplifications are not obligatory. If model is not simplified then for exact calculation of drain current numerical methods have to be used.

The second problem is a model obligation in a context of MOSFET modes of operation. Proposed model of current increment $\Delta{I_D}$ is obliging in all modes of MOSFET operation in the same degree as chosen model $I_{D0}$ (model of current without stress). On the other hand, current sensitivity on stress is dependent on mode of MOSFET operation. Significant factor in formula of current increment (\ref{Eq29}) is conductivity $\kappa$. The value of $\kappa$ (which is normalized output conductance $G_{DS}={\partial{I_{D0}}}/{\partial{V_{DS}}}$) in saturation region is small. It means current sensitivity on stress is smaller in saturation region than in linear region.

As a third problem there appears model universality. Described model of stress influence on MOSFET drain current was established with assumption that there is a transistor structure with a silicon channel. Matrix $\Pi$ (\ref{Eq21}) is used for description of piezoconductivity of silicon (which has diamond cubic crystal structure) in coordinate system conformable with elementary silicon crystal cell. On the other hand, instead of silicon the other material can be considered for transistor channel forming. For instance, it can be SiC (silicon carbide), which has zincblende (cubic) or hexagonal crystal structure \cite{Gut2009}. In this case, the shape of matrix $\Pi$ should be adjusted to actual crystal structure.

At the end, the weakness of the model should also be considered. This limitation is a nonlinearity of piezoconductivity coefficients versus stress level. Up to $100$MPa, assumption about linearity of the piezoconductivity phenomenon can be used. For stresses up to $200$MPa, non-linearity ranges from several to tens of percent \cite{Matsud1993}. Since magnitude of stress in MOSFETS is on $1$GPa range of level, assumption about linear piezoconductance cannot be applied. 

\section{Conclusions}

In this paper piezoconductivity phenomenon in MOSFET channel was discussed and model of drain current with possibility of stress consideration was proposed. General model which is obliging in all modes of MOSFET operation was simplified for practical use.

The analysis of proposed model combined with examination of stress components inherent in the channel as well as distributions of specific piezoconductance coefficients on a plane of channel showed which directions of transistor channel are desirable for improvement of MOSFET performances. This model gives possibility to predict optimum orientation of transistor channel, for the given stress state.

Described model was established with assumption that there is a transistor structure with a silicon channel. On the other hand, instead of silicon the other material can be considered for transistor channel forming. For instance, it can be SiC (silicon carbide), which has zinc blende (cubic) or hexagonal crystal structure. In this case, the shape of piezoconductivity matrix should be adjusted to actual crystal structure. By this means, in this paper general approach to modeling of stressed MOSFET is proposed.

\bibliography{mojeMOSFET}
\bibliographystyle{unsrt}
\end{document}